\documentclass[aps,prl,twocolumn,showpacs,superscriptaddress,amsmath,amssymb]{revtex4}

\usepackage{graphicx} 

\begin{document}

\title{Bulk Spin Polarization of Co$_{(1-x)}$Fe$_x$S$_2$}

\author{C. Utfeld} 
\affiliation{H.~H.~Wills Physics Laboratory, University of Bristol, Tyndall Avenue, 
Bristol BS8 1TL, United Kingdom} 
\author{S.~R. Giblin}
\affiliation{ISIS Facility, Rutherford Appleton Laboratory, Chilton, Oxfordshire
OX11 0QX, United Kingdom} 
\author{J.~W. Taylor} 
\affiliation{ISIS Facility, Rutherford Appleton Laboratory, Chilton, Oxfordshire 
OX11 0QX, United Kingdom}
\author{J.~A. Duffy} 
\affiliation{Department of Physics, University of Warwick, Coventry CV4 7AL, United Kingdom} 
\author{ C. Shenton-Taylor}
\affiliation{Department of Physics, University of Warwick, Coventry CV4 7AL, United Kingdom} 
\author{J. Laverock} 
\affiliation{H.~H.~Wills Physics Laboratory, University of Bristol, Tyndall Avenue, 
Bristol BS8 1TL, United Kingdom} 
\author{S.~B. Dugdale} 
\affiliation{H.~H.~Wills Physics Laboratory, University of Bristol, Tyndall Avenue, 
Bristol BS8 1TL, United Kingdom}
\author{M. Manno} 
\affiliation{Department of Chemical Engineering and Materials Science, $421$ Washington 
Ave SE, University of Minnesota, Minneapolis $55455$, USA} 
\author{C. Leighton} 
\affiliation{Department of Chemical Engineering and Materials Science, $421$ Washington 
Ave SE, University of Minnesota, Minneapolis $55455$, USA} 
\author{M. Itou} 
\affiliation{Japan Synchrotron Radiation Research Institute, SPring-8, 1-1-1 Kouto, Sayo, 
Hyogo 679-5198, Japan} 
\author{Y. Sakurai} 
\affiliation{Japan Synchrotron Radiation Research Institute, SPring-8, 1-1-1 Kouto, Sayo, 
Hyogo 679-5198, Japan}

\date{\today}

\begin{abstract}
We report on a new method to determine the degree of bulk spin
polarization in single crystal Co$_{(1-x)}$Fe$_x$S$_2$ by modeling magnetic
Compton scattering with {\it ab initio} calculations. Spin-dependent Compton
profiles were measured for CoS$_2$ and Co$_{0.9}$Fe$_{0.1}$S$_2$. The {\it ab initio}
calculations were then refined by rigidly shifting the bands to provide the best
fit between the calculated and experimental directional profiles for each
sample. The bulk spin polarizations, $P$, corresponding to the spin-polarized
density of states at the Fermi level, were then extracted from the {\it refined}
calculations. The values were found to be $P=-72 \pm 6 \%$ and $P=18 \pm 7\%$
for CoS$_2$ and Co$_{0.9}$Fe$_{0.1}$S$_2$ respectively. Furthermore,
determinations of $P$ weighted by the Fermi velocity ($v_F$ or $v_F^2$) were
obtained, permitting a rigorous comparison with other experimental data and
highlighting the experimental dependence of $P$ on $v_F$.
\end{abstract}

\maketitle

Spintronic materials are the subject of considerable research and rapid
technological development. In these materials and devices, the goal is to
exploit the electron spin for applications such as data storage and read-heads.
Some commercial devices exist, for example magnetic random access memory devices (MRAM) being
based on the concept of spin-dependent tunneling magnetoresistance
\cite{wolf2001}. Because of the potential for the development of novel devices
using spin transport in semiconductors, so-called spin-injector materials (which
would act as a source of highly spin-polarized electrons) currently attract
considerable interest. An obvious choice for a spin injector is a ferromagnetic
system since it naturally has an imbalance of the electron spin population at the
Fermi level and thus possesses a degree of spin polarization. Ideally, materials
for applications would be fully spin-polarized `half-metals' (HMs), where the
density of states (DOS) at the Fermi energy ($E_{F}$) is finite for one spin but zero for
the other, such that carriers of only one spin exist at the Fermi level.
However, ferromagnets typically only possess partial spin polarization. Despite
much theoretical and experimental effort, only a few promising candidates, such
as CrO$_2$ \cite{ji2001,bigJeff}, optimally doped La$_{0.7}$Sr$_{0.3}$MnO$_3$
\cite{coey} and Co$_{2}$MnAl$_{(1-x)}$Sn$_x$ \cite{co2mnal2,co2mnal} have been
found to exhibit high polarizations. Recently, Co$_{(1-x)}$Fe$_x$S$_2$ has
garnered interest resulting from predictions of half-metallicity and the
potential to {\it tune} the polarization via Fe doping. The consensus is that
doping simply adjusts the Fermi level
\cite{zhao1993,mazin2000,wang2005,leighton2007,wang2006a}, without altering the
bandstructure. The possibility of altering the magnitude of the polarization
makes this material ideal for fundamental research on spin
polarization.

The spin polarization can be defined simply in terms of the
spin-dependent density of states (DOS), $N_{\uparrow/\downarrow}$, at the Fermi
level. However, in order to facilitate comparison with various experimental
techniques, it is useful to expand this definition to weight
$N_{\uparrow/\downarrow}$ with the Fermi velocity, ${v_{F,
\uparrow/\downarrow}}$. Hence, following Mazin
\cite{mazinPRL}, {\it P$_n$} can be defined such that, 

\begin{equation} 
P_n = \frac{N_{\uparrow} v_{F, \uparrow}^n- N_{\downarrow} 
v_{F,\downarrow}^n}{N_{\uparrow} v_{F, \uparrow}^n + N_{\downarrow}v_{F,
\downarrow}^n}.  
\end{equation}

For $n=0$, as quoted by most theoretical studies, $P_0$ is solely defined
by the DOS. Weighting with the Fermi velocity accounts for transport properties,
with $n=1$ in the ballistic or $n=2$ in the diffusive regimes. Experimentally
the direct measurement of the polarization has proven to be a challenge. In
particular, except for the case of a HM where $P_n$ is $100\%$,
the appropriate value of $n$ is often ambiguous, and consequently comparison with
theory is possible, but notoriously difficult \cite{zhang2008, mazinPRL,mazin2004,mueller09}. 
In point-contact Andreev reflection (PCAR) measurements, for example, the value of $n$ is 
not the same for all materials \cite{mazin2004}. Bulk techniques such as saturation 
magnetization or transport measurements can only give an indication of half-metallicity, and
techniques which give access to the magnitude of $P$ directly such as spin
polarized photoemission or PCAR are heavily dependent on the surface quality of
the sample.

In this Letter, we demonstrate that a magnetic Compton scattering study combined with 
{\it ab initio} electronic structure calculations can be used to extract $P_n$ 
in the representative pyrite-type series Co$_{(1-x)}$Fe$_x$S$_2$. Fine tuning the theory 
to experiment enables the spin polarization to be determined for $n=0$ as well as the 
weighted values for $n=1,2$.  

Magnetic Compton scattering is a valuable tool for the
investigation of spin-dependent effects as the incoming hard x-rays directly
probe the twice-integrated momentum distribution of the unpaired electrons in
the bulk \cite{cooper1985}. A Compton profile (CP) represents the 1D projection
of the electron momentum density. If the difference between the CPs is measured for opposite
applied magnetic field directions, the resulting difference profile
contains only the spin-dependent part of the momentum distribution; this is
referred to as the Magnetic Compton Profile (MCP). 

The directional MCPs can be
directly compared to the spin-dependent
momentum distributions which have been computed from the {\it ab initio}
electronic structure calculation (from which the
the polarization may also be calculated). It has been demonstrated previously
\cite{zs2006,zs2004} that the agreement with experiment may be refined by
iterating a rigid shift of the bands (with respect to the
Fermi level) contributing to the spin moment. In the
case described here, this optimization is performed by minimizing the
${\chi^{2}}$ calculated simultaneously between the computed and measured MCPs along
four crystallographic directions.
The bulk polarization $P$, and the weighted
polarization $P_n$ can then be calculated directly from the refined model bandstructure 
giving a direct experimental insight into not only the polarization
but also the effect of the Fermi velocity on $P$. The ability to extract all
possible permutations of the polarization allows simple comparison with results
from other studies.

\begin{figure}
\includegraphics[width=0.85\linewidth]{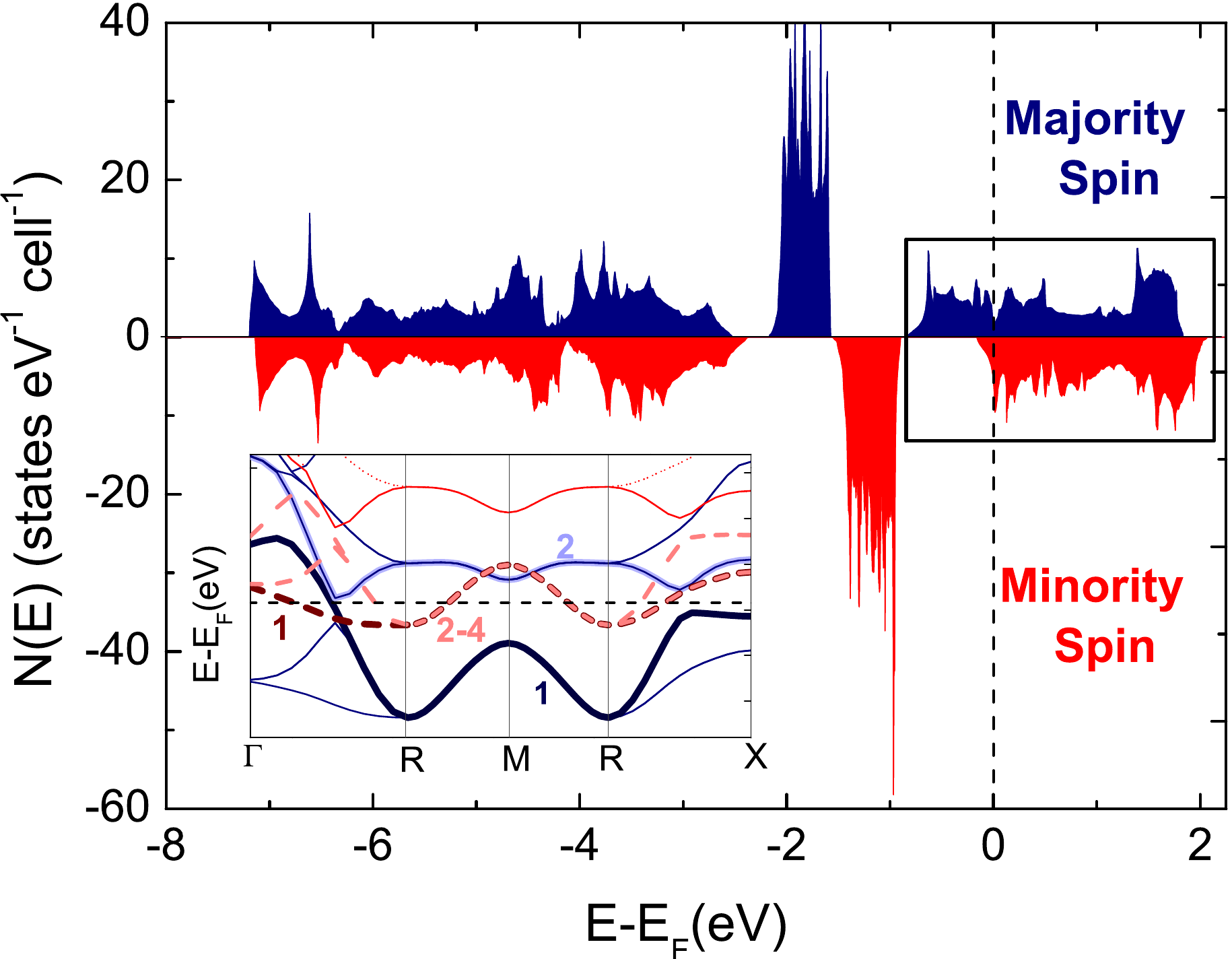}
\caption{\label{fig:dosbands} (Color online) The DOS of CoS$_2$, where the
sulphur $p$- and cobalt $d$-manifold are depicted for the majority and minority 
 spin states. The inset shows the bandstructure of the $e_g$-manifold around 
the Fermi level along a high symmetry path in the cubic Brillouin Zone. Overall six
bands composed of two in the majority (blue, solid lines) and four in the minority 
(red, dashed lines) channel cross the Fermi level.}
\end{figure}

In order to test our method, we calculated the spin polarization of Ni. For Ni,
a transport measurement determined $P_2$ to be $23\pm3\%$ experimentally
\cite{ni_p_data}, whereas {\it ab initio} calculations predict $P_2=0\%$
\cite{mazinPRL}. For our refinement, we used experimental MCPs for four
crystallographic directions from Dixon {\it et al.} \cite{dixon1998}. The refinement increased the value of $P_2$ from our {\it ab initio} 
value of 0\% to $20\pm2\%$, which agrees well with the previous experimental
result. This simple test shows that it is a robust method for determining $P_n$ 

\begin{figure}[b]
\includegraphics[width=1\linewidth]{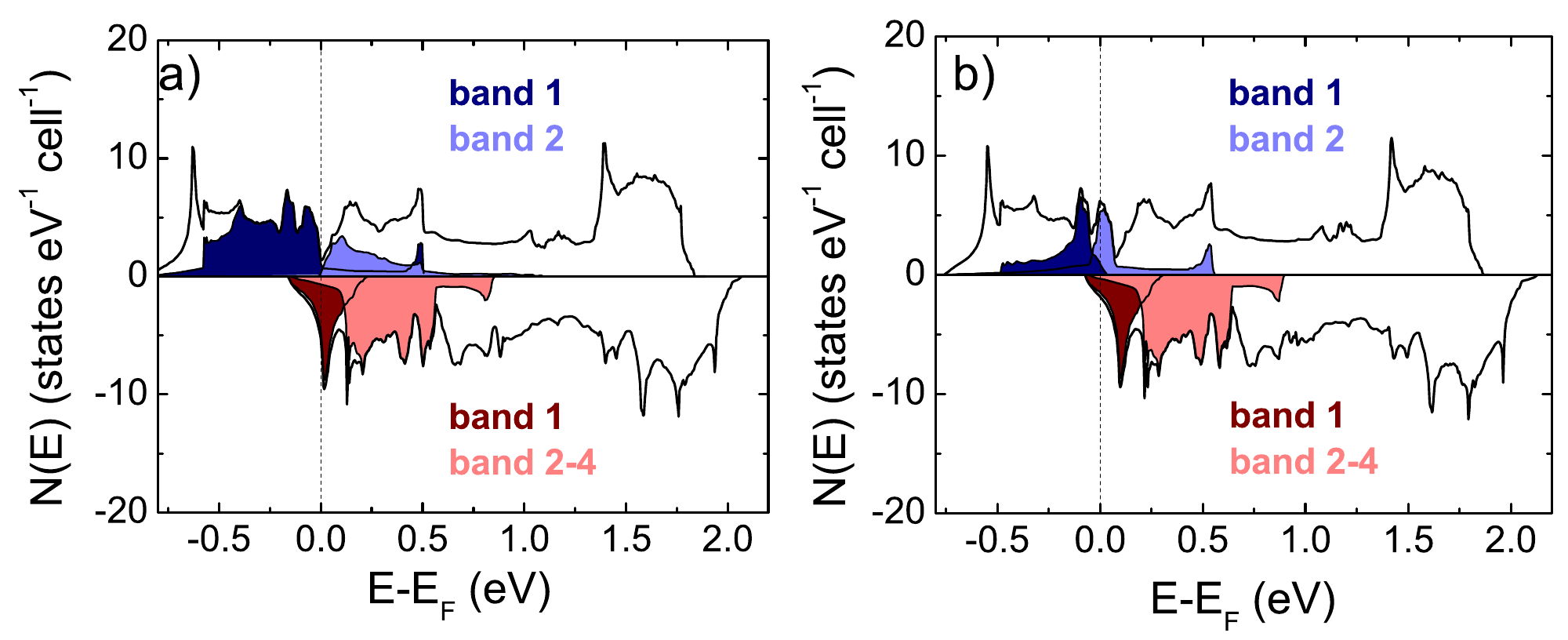}
\caption{\label{fig:partialDOS} (Color online) The DOS in the $e_g$-manifold of
(a) CoS$_2$ and (b) Co$_{0.9}$Fe$_{0.1}$S$_2$ is composed of six bands around
the Fermi level. As in Fig. \ref{fig:dosbands} the DOS of the
majority and minority bands are blue and red, respectively. The dark and light colors signify
band 1 and band 2 (or in the minority channel, the combination of bands 2--4), respectively.}
\end{figure}

\begin{figure*}[htp]
\includegraphics[width=0.48\linewidth]{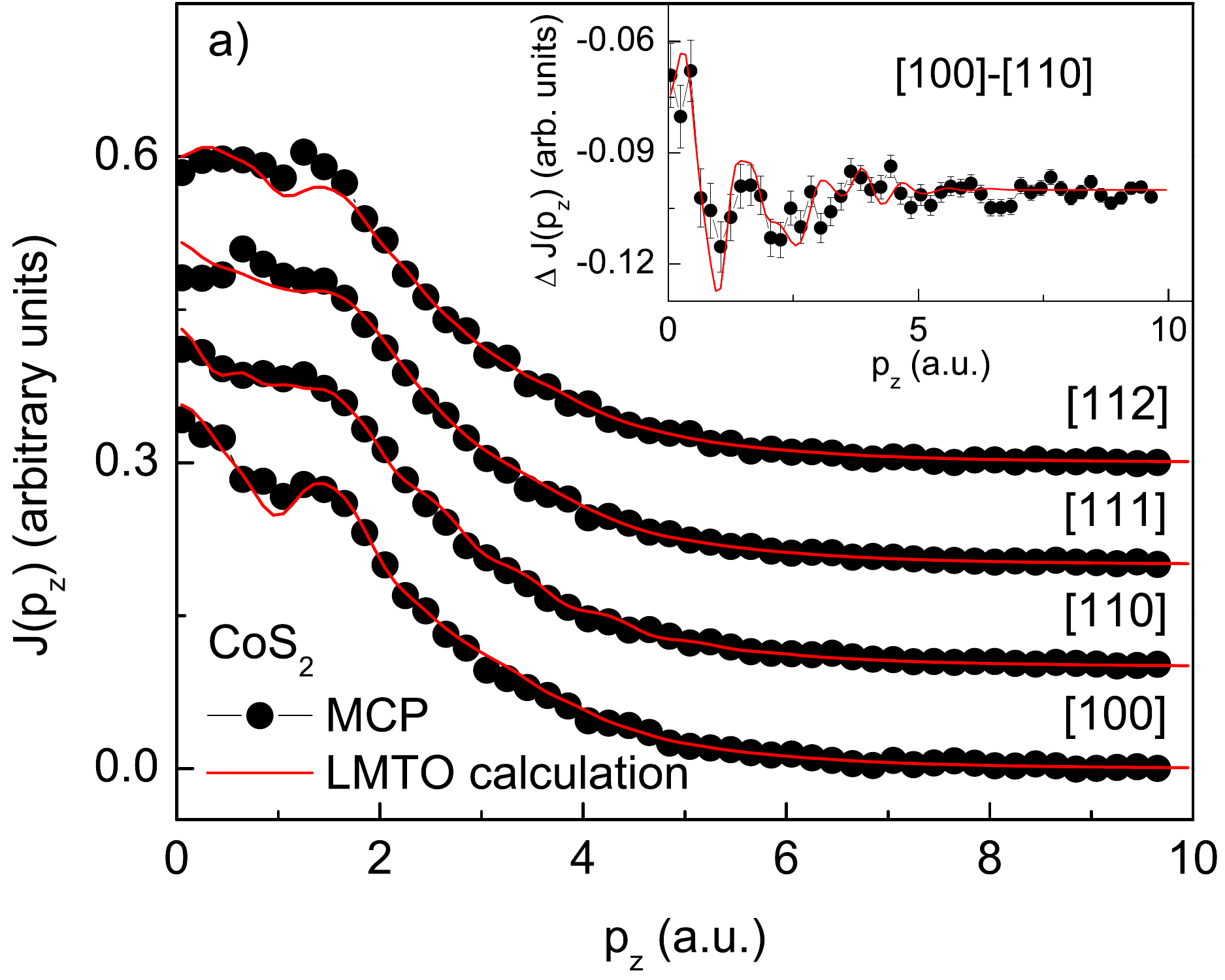}
\hspace{0.2in} \includegraphics[width=0.48\linewidth]{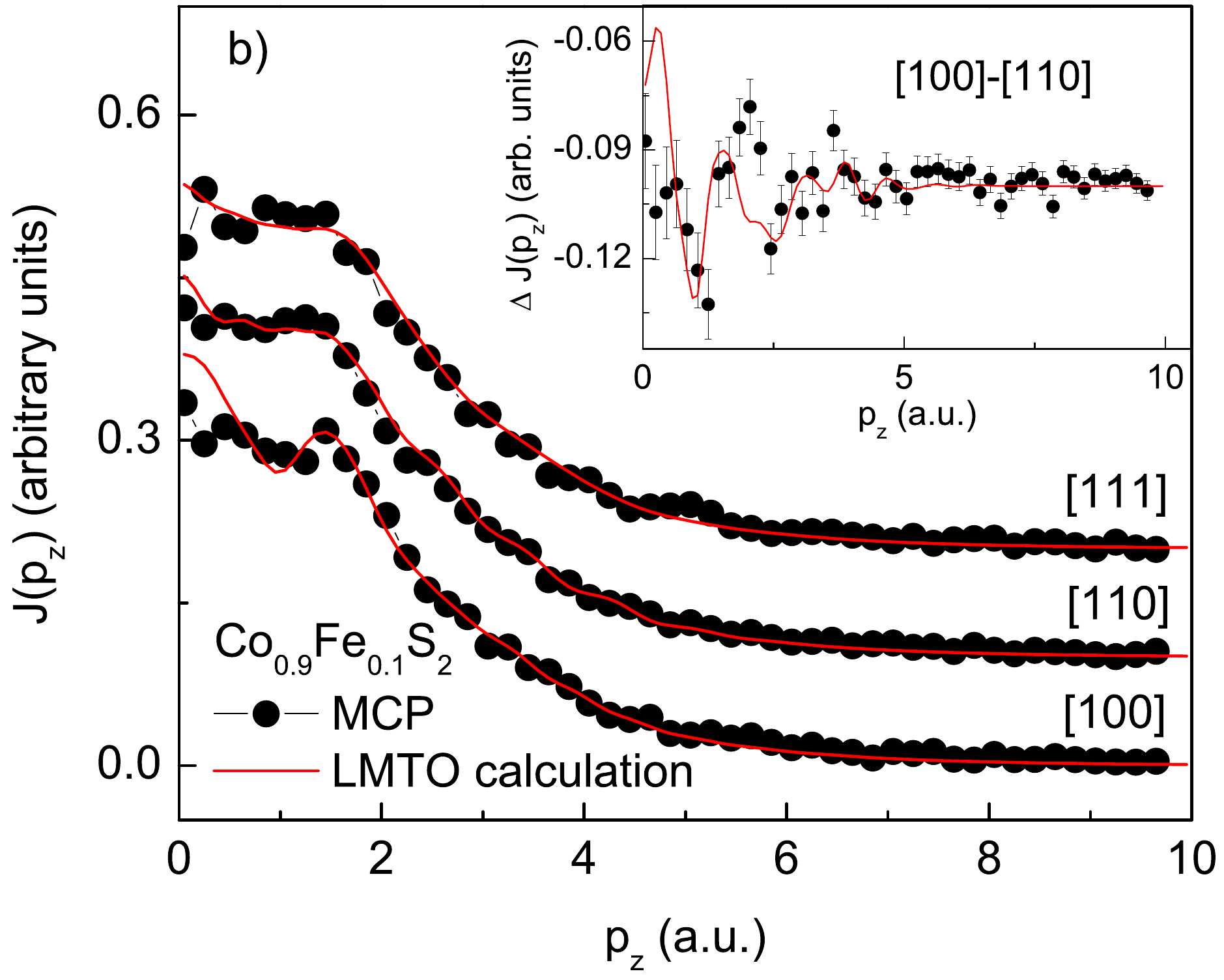}
\caption{\label{fig:mcps}(Color online) Experimental and LMTO MCPs for (a)
CoS$_2$ and (b) Co$_{0.9}$Fe$_{0.1}$S$_2$ resolved along different
crystallographic directions. For clarity the profiles are offset by 0.2 and
statistical errors are smaller than the symbol size. The insets of both graphs
show the directional differences for each composition between the [100] and
[110].}
\end{figure*}

Bandstructure calculations predict CoS$_2$ to be a highly
spin-polarized ferromagnet, just missing half-metallic behavior \cite{kwon2000}.
The Fermi level lies low in the conduction bands and doping with the
isostructural semiconductor FeS$_2$ results in a reduction in the occupation of
one spin channel. It has been predicted that this would enable the polarization
to be tuned to half metallicity \cite{zhao1993,mazin2000,wang2005}.
Saturation magnetization and transport measurements indicate a high
spin polarization hinting towards possible half-metallicity at doping levels
between $7-10\%$. The trend in the total DOS at $E_F$ extracted from heat
capacity and nuclear magnetic resonance follow this conjecture
\cite{wang2006a,kuhns2006}. However, PCAR
measurements of the absolute value of the magnetization reveal magnitudes of up
to $\lvert P \rvert=64\%$ in the pure (single crystal, \cite{wang2006b}) and
$\lvert P \rvert=85\%$ for $x=0.15$ (polycrystalline, \cite{wang2005}) in the
doped system. Complementing these results, anisotropic magnetoresistance
measurements show a change in sign of the polarization on doping, as predicted
by {\it ab initio} calculations \cite{wang2005,umemoto2006}. Although the
qualitative behavior of both experiment and theory are similar, the absolute
values of $P$ disagree \cite{note}. This may be attributed in part to the sensitivity of
PCAR to $v_F$ (such that $n>0$), as opposed to the theoretical values which
refer to $P_0$.

For the current study, measurements were performed on CoS$_2$ and
Co$_{0.9}$Fe$_{0.1}$S$_2$ single crystals to investigate polarization effects at
low doping. High quality single crystals were prepared by the chemical vapor
transport method \cite{wang2006a}. Particular care was taken to
ensure stoichiometry and to avoid sulphur deficiency, which is known to affect
the measured polarization \cite{leighton2007,cheng2003}. The magnetic Compton
scattering experiments were carried out at the
beamline (BL08W) of SPring-8, Japan. All experimental data were 
collected at a temperature of 10K. The MCPs were subsequently corrected for 
detector efficiency, absorption, the relativistic cross-section and multiple scattering 
effects. The value of the saturation magnetization measured with a SQUID at 
10K was determined to be 0.83 $\mu_B/$Co and 1.04 $\mu_B/$Co for the pure 
and $x=0.1$ samples.

The electronic stucture of CoS$_2$ was computed using the linear muffin-tin
orbital method (LMTO) within the atomic sphere approximation including combined
correction terms \cite{barbielini2003}. The exchange correlation part of the
potential was described within the local spin density approximation (LSDA) and
the effect of doping was incorporated via the virtual crystal approximation
(VCA). As observed in previous calculations, the moment at the experimental
lattice constant of 5.5\,\AA\ is substantially underestimated
\cite{yamada1998,ogura2007} and so the calculations presented here have been
performed at 5.71\,\AA, producing moments which are 0.85\,$\mu_B/$Co and
0.92\,$\mu_B/$Co for the pure and $x=0.1$ systems respectively, in good
agreement with the experimental values.

The results of the calculation for CoS$_2$ are depicted in Fig.
\ref{fig:dosbands}. The spin-resolved DOS show the sulphur $2p$-states below the
cobalt $3d$ manifolds. These states, which are hybridized with sulphur $2p$-states,
are split by the crystal field into $t_{2g}$- and $e_g$-manifolds. The Fermi
level lies low in the $e_g$-manifold where two majority bands and four minority
bands cross $E_F$ (see inset of Fig. \ref{fig:dosbands}). The partial DOS
contributed by each band are depicted in Fig. \ref{fig:partialDOS}(a). In the undoped
compound, in both the majority and minority spin channel, band 1 holds the largest
DOS and is therefore expected to have the biggest impact on $P_n$. The
calculation predicts a polarization $P_0 = - 58 \%$. The bandstructure and
partial DOS for $x=0.1$ are broadly similar to the pure case and lead to $P_0 =
40 \%$ (see Fig. \ref{fig:partialDOS}b). Our calculations predict
half-metallicity to occur for $x\geq0.3$. Overall these results are consistent
with other theoretical band predictions
\cite{umemoto2006,ogura2007,ramshea2004,yamada1998}.

The experimental and calculated MCPs, resolved along different crystallographic
directions, are shown in Fig.\ \ref{fig:mcps}(a) and \ref{fig:mcps}(b) for
CoS$_2$ and Co$_{0.9}$Fe$_{0.1}$S$_2$. The {\it ab initio} theoretical results
are in excellent agreement with the experimental MCPs. The characteristic
structures in the directional differences of the magnetic Compton profiles serve
as a rigorous check on the calculation, and these are shown for the pairs of
directions [100] - [110] in the insets of Fig. \ref{fig:mcps}. In order to
optimize the agreement between theory and experiment the energy bands were
refined, as described earlier, to provide the best fit to the experimental MCPs. Small energy shifts of
the bands crossing E$_F$ change the detailed shape of the resultant MCP, the
band positions were refined to give the best fit between the theoretical and
experimental MCPs, simultaneously for four crystallographic directions for
CoS$_2$ and three for Co$_{0.9}$Fe$_{0.1}$S$_2$. The spin moments calculated
from these fits agree well with the experimental values, providing an
independent consistency check.

Despite small energy shifts of the bands ($20-60$ meV), the fitting procedure sees the
$\chi^2$ parameter reducing by $30-50\%$ from its starting value to the minimum
and has an appreciable impact on the polarization. In the undoped case, the polarization
decreases from the original value of $P_0=-58\%$ to $P_0=-72 (\pm 6) \%$. Here
the error (which is generally larger for more dispersive bands) reflects the
uncertainty of the fit. This decrease can be mainly attributed 
to majority band 2 being shifted above E$_F$ and the DOS of band 1 being 
decreased. For the $x=0.1$ compound, the polarization decreases from the 
original value of $P_0=40\%$ to $P_0=18\% (\pm 7)$, which is primarily due 
to the lowering of majority band 2 and an increased DOS for the minority band.

\begin{figure}[t] \includegraphics[width=0.85\linewidth]{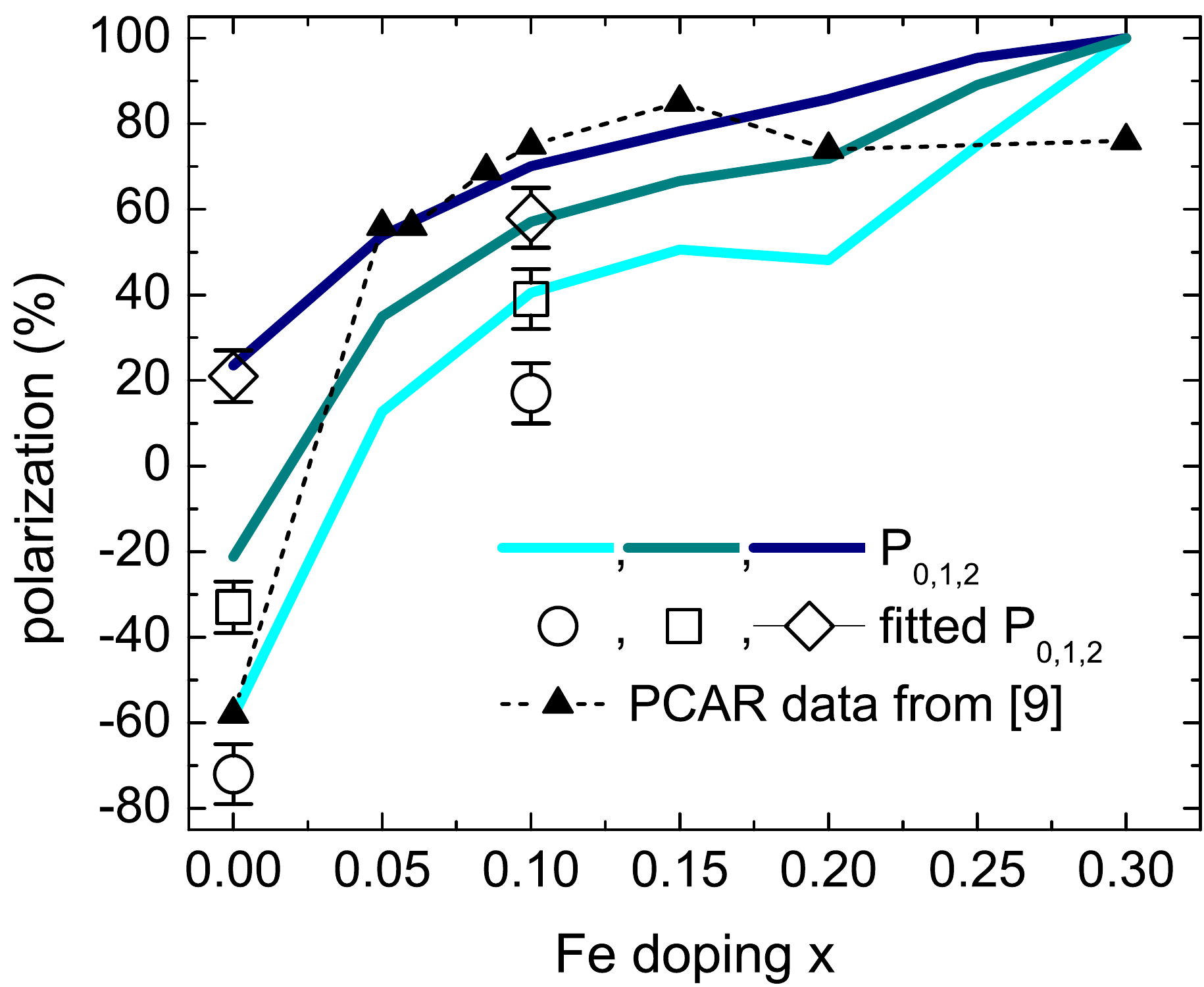}
\caption{\label{fig4} (Color online) Experimental PCAR (filled triangles)
polarization data compared to $P_{0,1,2}$ from the LMTO calculation (lines) as a
function of composition in 5\% steps. The tuned calculations are also shown,
where the bare $P$ is also weighted by the Fermi velocity (open symbols). Note
that the sign change in the PCAR data has been inferred from AMR measurements.}
\end{figure}

The fitted bulk polarization $P_{n}$ obtained for CoS$_2$ and
Co$_{0.9}$Fe$_{0.1}$S$_2$ are depicted in Fig. \ref{fig4} along with the results
from the unfitted calculation for different compositions. The results of the
original calculation for $P_0$ at different $x$ are in good agreement with
previous theoretical studies \cite{umemoto2006}. However, the weighting with the
Fermi velocity has a significant effect on the nominal value of the
polarization, particularly in the pure system where a weighting with $v_F^2$
results in a positive polarization. The evolution of the fitted $P_n$ is
qualitatively consistent with indirect measurements such as saturation
magnetization, but imply that the maximum $P$ is not obtained at $x=0.1$.

Our results demonstrate the tunability of the different polarization functions
across the series of Co$_{(1-x)}$Fe$_x$S$_2$. The data confirm that $P_0$ is
negative in the pure and positive for the doped compound. It is instructive to
compare our values for $P_n$ with the PCAR results. In the case of the pure
compound the value of $P_{\mbox{\footnotesize{PCAR}}}=\lvert 64\% \rvert$ is
between $P_0$ and $P_1$. However, for the $x=0.1$ composition
$P_{\mbox{\footnotesize{PCAR}}}$ is in close proximity to $P_2$ found in our
study. This could indicate a doping dependent crossover from the ballistic to
the diffusive regime in PCAR measurements, which is consistent with the increase
in resistivity observed for higher Fe-doping \cite{wang2006a}. Alternatively,
these discrepancies could be due to surface effects playing a role in the PCAR
measurements as pointed out by Leighton {\it et al.} \cite{leighton2007}.
Although PCAR has a penetration depth corresponding to the coherence length of
the superconductor, the electrons crossing the interfacial barrier experience
the surface potential. Studies on the surface state of CoS$_2$ show that sulphur
degradation could influence polarization measurements \cite{wu2008,xu2007}.
However, so far the surface stoichiometry of the alloyed composites has not been
investigated.

In summary, we present a new method to determine the bulk spin polarization.
Although harder to put into practice, this novel technique calculates $P_n$ from a 
refined model bandstructure that is fitted to experimental MCPs. Moreover, it is free 
from surface effects. The case of Ni shows that this approach is a
rigorous method of determining the bulk $P_n$. More importantly, the results on
CoS$_2$ and Co$_{0.9}$Fe$_{0.1}$S$_2$ illustrate the applicability of the
approach in a more complex system, and highlight the degree to which the
polarization is dependent on ${v_{F}}$. We speculate that there is a crossover
from the ballistic to the diffusive transport regimes on doping with Fe 
on the basis of the comparison with PCAR measurements. The
new method can be used to determine $P_n$ in such a way that its dependence on
${v_{F}}$ is unambiguous, and our results for Co$_{1-x}$Fe$_{x}$S$_2$
demonstrate the importance of this.

\section*{Acknowledgments} The authors thank II Mazin for helpful suggestions.
This experiment was performed with the approval of JASRI (Proposal No. 2006B1363). Work at UMN was
supported by NSF MRSEC under Grants No. DMR-0212302 and No. DMR-0819885.


\begin{thebibliography}{99}
\bibitem{note} Note: If we use $P$ instead of $P_n$ no regime can be
assigned.
\end{thebibliography}
\end{document}